\pdfoutput=1
\documentclass[preprint,sort&compress,12pt]{elsarticle}

\usepackage{units}
\usepackage{booktabs}
\usepackage{amsfonts}
\usepackage{amssymb}
\usepackage{amsmath}
\usepackage{epstopdf}
\usepackage{algorithm}
\usepackage{algpseudocode}
\usepackage{color}

\journal{Journal of Computational Physics}

\begin{document}

\begin{frontmatter}



\title{Accelerating a hybrid continuum-atomistic fluidic model with on-the-fly machine learning}


\author[ad1]{David Stephenson \corref{ca}}
\ead{david.stephenson@warwick.ac.uk}
\author[ad2]{James R. Kermode}
\author[ad1]{Duncan A. Lockerby}

\address[ad1]{School of Engineering, University of Warwick, Coventry CV4 7AL, UK}
\address[ad2]{Warwick Centre for Predictive Modelling, School of Engineering, University of Warwick, Coventry CV4 7AL, UK}

\cortext[ca]{Corresponding author}

\begin{abstract}
We present a hybrid continuum-atomistic scheme which combines molecular dynamics (MD) simulations with on-the-fly machine learning techniques for the accurate and efficient prediction of multiscale fluidic systems. By using a Gaussian process as a surrogate model for the computationally expensive MD simulations, we use Bayesian inference to predict the system behaviour at the atomistic scale, purely by consideration of the macroscopic inputs and outputs. Whenever the uncertainty of this prediction is greater than a predetermined acceptable threshold, a new MD simulation is performed to continually augment the database, which is never required to be complete. This provides a substantial enhancement to the current generation of hybrid methods, which often require many similar atomistic simulations to be performed, discarding information after it is used once.

We apply our hybrid scheme to nano-confined unsteady flow through a high-aspect-ratio converging-diverging channel, and make comparisons between the new scheme and full MD simulations for a range of uncertainty thresholds and initial databases. For low thresholds, our hybrid solution is highly accurate\,---\,within the thermal noise of a full MD simulation. As the uncertainty threshold is raised, the accuracy of our scheme decreases and the computational speed-up increases (relative to a full MD simulation), enabling the compromise between precision and efficiency to be tuned. The speed-up of our hybrid solution ranges from an order of magnitude, with no initial database, to cases where an extensive initial database ensures no new MD simulations are required.
\end{abstract}

\begin{keyword}
Multiscale modelling \sep Machine Learning \sep Hybrid methods \sep Micro/nanofluidics \sep Molecular Dynamics


\end{keyword}

\end{frontmatter}

\section{Introduction} \label{secIntro}
\noindent Almost all fluid engineering systems are multiscale in their nature. At the smallest scale, the fluid and surrounding environment are comprised of atoms, with interactions occurring across nanometers ($10^{-9}$ m) and over femtoseconds ($10^{-15}$ s), while the fluid flow is characterised by the scale of the system geometry, which is often many orders of magnitude larger. In most instances, the separation of scales is so large that the atomistic behaviour can be accurately incorporated into a continuum fluid description through empirical boundary conditions (e.g. no-slip at the walls) and constitutive relations (e.g. viscosity in the shear stress\,--\,strain rate relation). However, as some characteristic dimension of the system approaches the nanoscale (and microscale in gases), these approximations break down, and the fluid flow becomes highly dependent on atomistic phenomena \cite{surfaceEffects,contactLine,dropletBreakup,gasNoSlip}.

There are numerous applications in fluid dynamics where atomistic information is required to capture non-continuum/non-equilibrium phenomena, but the macroscopic flow develops over much larger length and time scales: for example, pumping technology that exploits thermal creep in a rarefied gas \cite{thermalCreep}; and low-power, high-throughput nanotube membranes for salt water desalination \cite{desalination}. The multiscale nature of these systems leads to a dual requirement for capturing the local atomistic-scale interactions and macro-scale fluid response. The complexity of the flow necessitates modelling with atomistic resolution, but the state-of-the-art techniques (molecular dynamics (MD) for dense fluid flows \cite{MDref}, and the direct simulation Monte-Carlo method (DSMC) for rarefied gas flows \cite{DSMCref}) are extremely computationally expensive. This limits their application to small system sizes, typically $\mathcal{O}$(100 nm$^3$), and short simulation times,  typically $\mathcal{O}$(100 ns), rendering many important engineering problems intractable, and limiting possibilities for comparison with experiments. 

Hybrid methods provide a promising framework for simulating such systems  by combining continuum (macro) and atomistic (micro) solvers and exploiting scale-separation where it exists to obtain a high accuracy, yet efficient, solution. There has already been extensive research into hybrid approaches in fluid dynamics (see, for example, recent reviews \cite{hybrid1, hybrid2, hybrid3,hybrid4}), with most methodological variations concerning either a) the coupling strategy between the macro and micro models, or b) the boundary conditions that are imposed on each model. Coupling is usually performed via the exchange of state variable or flux information, to ensure consistency between the macro and micro models at locations where an atomic resolution is required. The information passed from one model is then applied as a boundary condition to locally constrain the other model.

Broadly speaking, the types of problems that hybrid methods are designed to model can be divided into three categories according to common features: Type A, Type B, or Type C. In Type A problems, a continuum model provides a reasonably accurate solution for much of the flow field, and an atomistic solver is only required for regions that contain non-continuum/non-equilibrium effects such as an isolated defect, a singularity, or regions near nanoscale surfaces. This is often referred to as domain decomposition because the macro and micro domains are coupled across spatial interfaces. Examples of Type A problems include the moving contact line problem \cite{contactLine} and resolving singularities that occur in the corners of lid-driven cavity flows \cite{corner}. In Type B problems, a continuum model spans the entire system domain, and an atomistic resolution is required everywhere to continually refine the continuum solution by providing local constitutive/boundary information. This is an embedded coupling approach, with the macro and micro domains occupying the same physical space; an example of a Type B problem is capturing the non-Newtonian flow of a complex nanofluid \cite{nonNewt}. In Type C problems, high-aspect-ratio flows result in non-continuum/non-equilibrium effects persisting over the entire (or majority of the) cross section of the fluid stream, while hydrodynamic properties vary slowly in the streamwise direction. The former must be captured by an atomistic solver, while the latter can be described by a continuum. This is also an embedded coupling approach, with examples including the flow through nanotube desalination membranes \cite{desalination} and air-layer lubrication of liquid journal bearings \cite{bearing}. However, despite much progress, a common flaw of hybrid methods is that, often, similar and repetitious atomistic simulations are performed\,---\,i.e. information is used once, then wastefully discarded, despite configuration conditions remaining similar.

In recent years, machine learning (ML) techniques have been used in an increasingly wide range of disciplines \cite{MacKay2003}. Prominent examples based on Bayesian inference include the calibration of computer models \cite{Kennedy2001} or the closely related problem of developing surrogate models that can be computed much more cheaply \cite{Rasmussen2006}. These approaches can help to quantify uncertainty; for example, the uncertainty induced in density functional theory (DFT) calculations from the choice of exchange-correlation functional \cite{Aldegunde2016}, or the uncertainty induced in nanoscale flow simulations from the choice of interatomic potentials \cite{Angelikopoulos2013}. Hand-calibrated force fields have been successfully replaced by models built automatically from quantum mechanical (QM) reference data, using neural networks \cite{Behler2007} or Gaussian process regression \cite{Bartok2010}. This has enabled accurate models to be produced for materials that were conventionally difficult to describe precisely with interatomic potentials \cite{Szlachta2014}. However, all such approaches are limited by the training data used to fit the ML model. In particular, they are not transferable to situations not envisaged at the time of construction.
 
More recently, a new class of methods for ML-based atomistic simulation have been proposed, which addresses this limitation by avoiding ever attempting to build a full description of the reference database (i.e. potential energy surface). Instead, they accelerate MD simulations by computing new reference data `on the fly' where and when it is needed \cite{Li2015,Botu2015}. The reference data is added to a growing database that is used to infer the information needed to propagate the dynamics, greatly reducing the computational cost of MD simulations while retaining the link to QM-accurate methods \cite{Caccin2015}. Similar ideas have been applied to develop neural network models for hybrid continuum-atomistic flow simulations \cite{Asproulis2013}. However, optimising the architecture of parametric neural networks, while avoiding over- or under-fitting, is a challenging task. Moreover, in \cite{Asproulis2013}, the criterion for computing new reference data is somewhat ad hoc, based on a fixed difference between parameters in input space. Here, we improve upon this idea by adopting a non-parametric approach to regression\,---\,using a Gaussian process to model the atomistic database in a hybrid continuum-atomistic scheme. The non-parametric model helps avoid over- or under-fitting to the data, and employing Bayesian inference for output predictions means that the criterion for obtaining new reference data can be based on a target threshold variance in output space, leading to a new robust approach to hybrid modelling in fluid dynamics, with close-to-optimal information efficiency.

\section{Methodology} \label{secMethod}
\noindent The specifics of our new scheme will depend on the exact hybrid method chosen, but the basic idea remains the same. Each time the continuum solver seeks to obtain atomistic information, rather than immediately performing a new simulation, we first predict the simulation output based on the similarity between the current system configuration and all configurations in a pre-existing database. A new atomistic simulation is only performed when the uncertainty of this prediction is above a predetermined threshold, i.e. the system configuration is not well represented in the database; otherwise the prediction is deemed sufficient. The data from new simulations is used to augment the existing database, whose completeness is never required. As this database grows, progressively fewer atomistic simulations are required, because `new' system configurations are more rarely encountered.

\begin{figure}[]
\centering
\includegraphics[width=\textwidth]{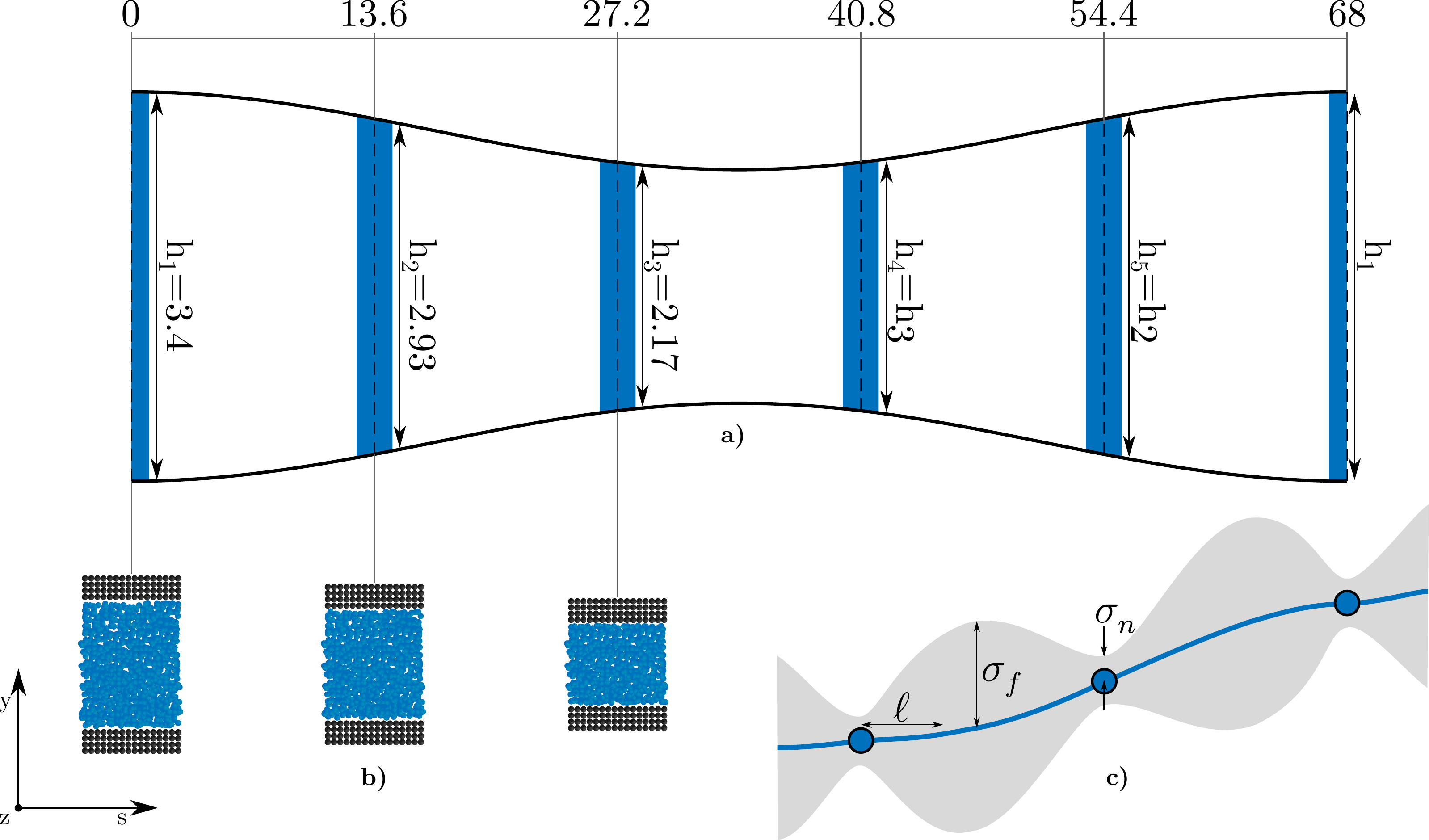}
\caption{Schematics of a) the multiscaled converging-diverging nanochannel, b) the micro subdomain decomposition, and c) the representation of the Gaussian-process predictive model, where the blue line is the mean and the grey envelope represents one standard deviation. All dimensions are in nm.}
\label{introFig}
\end{figure}

The system we will use as a benchmark for our new scheme is dense fluid flow through a converging-diverging channel, with the flow driven by a time-variant periodic external force $F_{\mathrm{ext}}(t)$. The geometry of the system is presented in Fig.\ \ref{introFig}a. We choose this as our benchmark system for two main reasons: 1) it is multiscale both spatially and temporally, demonstrating the potential for high computational savings; and 2) the results for the full atomistic simulation, recently published by Borg et. al. \cite{unsteadyIMM}, provide a useful basis for comparison. The hybrid method we use to model this system was developed by Borg et. al. \cite{unsteadyIMM}, and described in detail therein. However, for completeness, we provide a short summary below.

\subsection{Hybrid method} \label{secHybrid}
The benchmark system is a Type C problem, so non-continuum effects persist over the entire cross section, and spatial scale separation can only be exploited in the streamwise direction. As such, atomistic subdomain simulations must cover the entire channel height, and are placed at regular intervals in the streamwise ($s$-) direction according to the equation:
\begin{equation}
s_i = \frac{\left(i-1\right)}{N}L, \label{subsEq}
\end{equation}
where $i$ is the subdomain index, $N$ is the number of subdomains, and $L$ is the channel length. Note, the channel uses periodic boundary conditions (PBCs) in the ($s$-) direction, so the first subdomain is simultaneously located at the inlet and the outlet. The number of subdomains is set large enough to resolve the streamwise geometrical variation; here we, like Borg et. al., use $N=5$. The channel height $h(s)$ is a function of the streamwise position $s$:
\begin{equation}
h = a\left[\cos\left(\frac{2\pi s}{L}\right) - 1\right] + h_1, \label{heightEq}
\end{equation}
where $a=0.68$ nm is a constant and $h_1=3.4$ nm is the channel height at the inlet/outlet (see Fig.\ \ref{introFig}). Temporal scale separation exists because the characteristic time for the evolution of the macro model (e.g. the period of the external force) is much larger than the characteristic time for the development of the micro model (e.g. the start-up time from rest); therefore, each micro subdomain simulation is considered to be in a quasi-steady state.

The macro model consists of the unsteady one-dimensional equations for mass and momentum conservation. We use MD for the micro model, with atoms interacting through pairwise potentials and moving according to Newton's laws of motion (see \ref{appendA} for details). Coupling is performed by ensuring that the mass and momentum in each MD subdomain are consistent with the conservation laws of the macro model. For mass, we obtain
\begin{equation}
\frac{\partial \rho}{\partial t} + \left(\frac{1}{A}\right)\frac{\partial q}{\partial s} = 0, \label{continuityEq}
\end{equation}
where $\rho(s,t)$ is the density, $A$ is the cross-sectional area, and $q(s,t)$ is the mass flow rate which are time-averaged measurements in each MD subdomain simulation. For simplicity, the MD subdomains are simulated using PBCs in the $s$-direction (see Fig.\ \ref{introFig}b), which cannot support a pressure gradient. Therefore, for momentum flux to be hydrodynamically equivalent to that in the macro model, the total (external) force $F(s,t)$ applied to each atom is
\begin{equation}
F = F_{\mathrm{ext}} - \left(\frac{m}{\rho}\right)\frac{\mathrm{d}p}{\mathrm{d}s}, \label{momentumEq}
\end{equation}
where $m$ is the mass of a single atom.

\subsection{Machine learning} \label{secML}
\noindent For the machine learning model, we use Gaussian process (GP) regression. Here we provide only a brief overview of the approach, see Rasmussen and Williams \cite{Rasmussen2006} for further details. A non-parametric approach to regression is to model the unknown underlying function $f(\textbf{x})$ as random, with a GP distribution:
\begin{equation}
f(\textbf{x}) \sim \mathcal{GP}\big(\mu(\textbf{x}),K(\textbf{x},\textbf{x}')\big), \label{GPEq}
\end{equation}
where $\textbf{x}$ is a point in input space, $\mu(\textbf{x})$ is the mean function and $K(\textbf{x},\textbf{x}')$ is the covariance function for the process of interest, which models the covariances between function outputs $y$ and $y'$ at $\textbf{x}$ and $\textbf{x}'$, respectively. For our system, the input vector $\textbf{x} = \left\{h, \rho, F\right\}$ is composed of three macroscopic input variables: channel height $h$, density $\rho$, and force $F$. Here, we will use two separate GPs to predict the output for the mass flow rate $y=q$ and pressure\footnote{averaged across the channel cross section.} $y=p$ (note, for pressure, the input vector only depends on channel height and density).

A GP is a generalisation of a Gaussian probability distribution to an infinite number of dimensions. For each point in input space $\textbf{x}$, we model the output $y=f(\textbf{x})$ as a random variable with a univariate Gaussian distribution; similarly, we can model the output across the infinite range of input space using a GP. Starting from some prior belief in our function $P(f)$ (defined by our choice of covariance function, discussed below), we perform $M$ training MD simulations with inputs
\begin{equation}
X = 
\begin{bmatrix}
h_1 & h_2 & \dots & h_M \\
\rho_1 & \rho_2 & \dots & \rho_M \\
F_1 & F_2 & \dots & F_M
\end{bmatrix}, \label{trainingInputs}
\end{equation}
and observe outputs $\langle\textbf{y}\rangle$ (specifically, mass flow rate $\langle\textbf{q}\rangle$ and pressure $\langle\textbf{p}\rangle$). Using Bayesian inference, this leads to a posterior belief in the function $P(f\vert \langle\textbf{y}\rangle,X)$ given by
\begin{equation}
P(f\vert \langle\textbf{y}\rangle,X) = \frac{P(\langle\textbf{y}\rangle\vert X,f)P(f)}{P(\langle\textbf{y}\rangle\vert X)}, \label{bayesEq}
\end{equation}
where $P(\langle\textbf{y}\rangle\vert X,f)$ is the likelihood of observing the outputs, given the prior belief, and $P(\langle\textbf{y}\rangle\vert X)$ is the marginal likelihood. The angle brackets represent time-averaged measurements from MD simulations. The mean of the posterior $\mathbb{E}[f\vert \langle\textbf{y}\rangle,X]$ will closely resemble the measured outputs $\langle\textbf{y}\rangle$ near the input data points $X$, and the uncertainty (variance) will be low; however, the uncertainty away from the observed data remains large (see Fig.\ \ref{introFig}c). The posterior predictions thus become more accurate as the database grows and covers more of input space. As each data point `speaks', the GP can be considered to have a finite, but unbounded, number of parameters, which grow with the database.

To model the covariance between outputs, we use the squared exponential kernel
\begin{equation}
K(\textbf{x}_i,\textbf{x}_j) = \sigma_f^2 \exp\left(-\frac{d_{ij}^2}{2\ell^2}\right), \label{quadraticKernelEq}
\end{equation}
where $\sigma_f$ and $\ell$ are hyperparameters\,---\,with $\sigma_f^2$ representing the variance of the unknown function and $\ell^2$ representing the length scale\footnote{Note, this is a length scale in input space of distances $d_{ij}$, and should not be confused with any physical length scale used in the multiscale modelling.}\,---\,and $d_{ij}^2$ is the normalised Euclidean distance between the points $\textbf{x}_i = (h_i, \rho_i, F_i)$ and $\textbf{x}_j = (h_j, \rho_j, F_j)$ in input space: 
\begin{equation}
d_{ij}^2 = \left(\frac{h_i - h_j}{\overline{\Delta h}}\right)^2 + \left(\frac{\rho_i - \rho_j}{\overline{\Delta \rho}}\right)^2 + \left(\frac{F_i - F_j}{\overline{\Delta F}}\right)^2. \label{d2Eq}
\end{equation}
The denominator of each term in equation \eqref{d2Eq} is the mean separation between inputs in the database. This normalisation ensures that each input variable has approximately equal weight, and is recalculated when the covariance matrix is updated as the database grows. For the pressure kernel, only the first two terms in equation \eqref{d2Eq} are included.

When making predictions for the outputs, we include the inherent noise in the measured outputs from MD simulations by adding Gaussian noise with variance $\sigma_n^2$ to the covariance matrix
\begin{equation}
C(X,X) = K(X,X) + \sigma_n^2I, \label{covarianceEq}
\end{equation}
where $I$ is the identity matrix and $\sigma_n^2$ can be interpreted as an additional hyperparameter (see \S\ref{secHypers}).

To make predictions for a set of test configurations with inputs $X_*$, it can be shown that Eq.~\ref{bayesEq} leads to a posterior distribution for the outputs $\textbf{y}_*$ given by
\begin{equation}
\mathbf{y}_* | X, \mathbf{y}, X_* \sim\mathcal{N}\left( \bar{\textbf{y}}_*, \mathrm{cov}(\textbf{y}_*)\right), \label{posterior}
\end{equation}
where the mean is
\begin{equation}
\bar{\textbf{y}}_* = \mu(X_*) + K(X_*,X) C(X,X)^{-1}\left( \langle\textbf{y}\rangle - \mu(X) \right), \label{predMeanEq}
\end{equation}
and the covariance matrix
\begin{equation}
\mathrm{cov}(\mathbf{y}_*) = K(X_*,X_*) - K(X_*,X) C(X,X)^{-1}K(X,X_*). \label{predVarEq}
\end{equation}
The variances $\boldsymbol{\sigma}_{y_*}^2$ of the new predictions are the diagonal of the posterior covariance matrix $\mathrm{cov}(\mathbf{y}_*)$. In our scheme, whenever the predicted standard deviation $\sigma_{y_*}$ exceeds a pre-determined uncertainty threshold $\sigma_t$, the prediction is deemed insufficiently accurate and a new MD simulation is automatically performed and added to the teaching database.

\subsection{Hyperparameters} \label{secHypers}
Our non-parametric model nevertheless includes three hyperparameters $\boldsymbol\theta = (\sigma_f^2, \ell^2, \sigma_n^2)$, where $\sigma_f^2$ is the signal variance, i.e. the expected variance of the output function away from any known data points; $\sigma_n^2$ is the noise variance, i.e. the variance at known data points (see Fig.\ \ref{introFig}); and $\ell^2$ is the length scale, i.e. a measure of the distance between input points beyond which the outputs become uncorrelated; see Fig.\ \ref{introFig}c for a visual representation. The normalisation used here, in equation \eqref{d2Eq}, makes the choice of hyperparameters highly intuitive; it is likely that $\sigma_f^2$ and $\ell^2$ will be approximately unity, and $\sigma_n^2$ will be some fraction of $\sigma_f^2$. We verified this intuition by computing the marginal likelihood of observing the measured outputs $\langle\textbf{y}\rangle$ given the inputs $X$ and hyperparameters $\boldsymbol\theta$, given by
\begin{equation}
\log P(\langle\textbf{y}\rangle\vert X, \boldsymbol\theta) = -\frac{1}{2}\langle\textbf{y}^\top\rangle C(X,X)^{-1}\langle\textbf{y}\rangle - \frac{1}{2}\log\lvert C(X,X)\rvert - \frac{n}{2}\log 2\pi \label{evidenceEq}
\end{equation}
where it should be noted $C(X,X)$ depends on $\boldsymbol\theta$. We wish to maximise $\log P(\langle\textbf{y}\rangle\vert X, \boldsymbol\theta)$ with respect to $\boldsymbol\theta$, i.e. find the hyperparameters which maximise the likelihood of the outputs we observed. We calculated the log marginal likelihood over a small sample of training data points, which would later be used for initial databases (see Case D5 from Table \ref{preDatabases} in \S\ref{secResultsData}). We did not optimise for $\sigma_n^2$ as this could be estimated directly from the data, by calculating the standard deviation of the mean output for each MD simulation (using the raw time-instantaneous data) and averaging over the number of MD simulations. We found ${\sigma_n}_q=0.05$ $\mathrm{ng/s}$ for mass flow rate and ${\sigma_n}_p=0.003$ MPa for pressure\footnote{Note, we present the hyperparameters in terms of standard deviation, rather than variance, for the sake of clarity, as they share units with the outputs.}. Solving \ref{evidenceEq} numerically to maximise the log marginal likelihood for the other two hyperparameters, we obtained $\sigma_f=1$ ($\mathrm{ng/s}$ and MPa for mass flow rate and pressure, respectively) and $\ell=1$ for both mass flow rate and pressure, confirming our intuition.

\subsection{Implementation} \label{secAlg}
\noindent The procedure for our machine-learning-accelerated hybrid method is described in Algorithm~\ref{alg1}. After the initial databases for pressure and mass flow rate are loaded, and the number of subdomains $N$ is chosen, the location $\textbf{s}_{1...N}$ and height $\textbf{h}_{1...N}$ of the subdomains are calculated using equations \eqref{subsEq} and \eqref{heightEq}, respectively. The subdomains are initialised with densities $\boldsymbol{\rho}_{1...N}^1=\left\{1331, 1320, 1278, 1273, 1312\right\}$ kg/m$^3$, equivalent to the initial densities in the full MD simulation.

\begin{algorithm}
\caption{Hybrid scheme with on-the-fly machine learning}
\label{alg1}
\begin{algorithmic}[1]
\State load initial database $C_p(X_p,X_p)$, $X_p=\left\{\textbf{h},\boldsymbol{\rho}\right\}_p$, $\langle\textbf{p}\rangle$, $\boldsymbol{\theta}_p=\left\{\sigma_f^2,\ell^2,\sigma_n^2\right\}_p$
\State load initial database $C_q(X_q,X_q)$, $X_q=\left\{\textbf{h},\boldsymbol{\rho},\textbf{F}\right\}_q$, $\langle\textbf{q}\rangle$, $\boldsymbol{\theta}_q=\left\{\sigma_f^2,\ell^2,\sigma_n^2\right\}_q$
\State set case variables $N, \textbf{s}_{1...N}$, $\textbf{h}_{1...N}$, $\boldsymbol{\rho}_{1...N}^1$, $t_{\mathrm{end}}$, $\Delta t$, $\sigma_t$
\Procedure{time evolution}{}
    \For{$i\gets 1,t_{\mathrm{end}} / \Delta t$}
        \State $t\gets i\Delta t$
        \State $X_*\gets\left\{\textbf{h}_{1...N},\boldsymbol{\rho}_{1...N}^i\right\}$
        \State $\left\{\bar{\textbf{p}}_{1...N}^i, {\boldsymbol{\sigma}_p}_{1...N}^i\right\} \gets$ \Call{GPreg}{$C_p,X_p,\langle\textbf{p}\rangle,X_*;\boldsymbol{\theta}_p$}$\;\;\;\; \triangleright$ Eqs.\ \eqref{predMeanEq}, \eqref{predVarEq} 
        \State $\boldsymbol{p\prime}_{1...N}^i \gets \Delta\bar{\textbf{p}}_{1...N}^i/\Delta\textbf{S}_{1...N}$
        \State set external forcing $F_{\mathrm{ext}}^i$
        \For{$j\gets 1,N$}
            \State $F_j^i\gets F_{\mathrm{ext}}^i - (m/ \rho_j^i) {p\prime}_j^i\;\;\;\;  \triangleright$ Eq.\ \eqref{momentumEq}

            \State $\textbf{x}_*\gets\left\{h_j^i, \rho_j^i, F_j^i \right\}$
            \State $\left\{\bar{q}_j^i, {\sigma_q}_j^i\right\} \gets$ \Call{GPreg}{$C_q,X_q,\langle\textbf{q}\rangle,\textbf{x}_*;\boldsymbol{\theta}_q$}$\;\;\;\; \triangleright$ Eqs.\ \eqref{predMeanEq}, \eqref{predVarEq}
            \If{${\sigma_q}_j^i > \sigma_t$}
                \Procedure{Perform new MD simulation}{}
                    \State find nearest prior input $\textbf{x}_{\mathrm{in}}$
                    \State $t_s \gets$ \Call{pseudoMD}{$\textbf{x}_{\mathrm{in}},\textbf{x}_*$}
                    \State $\left\{\bar{q}_j^i, {\sigma_q}_j^i, \bar{p}_j^i, {\sigma_q}_j^i\right\} \gets$ \Call{MD}{$\textbf{x}_{\mathrm{in}}$, $\textbf{x}_*$, $t_s$}
                    \State append $X_q \gets \left[X_q\; \textbf{x}_*\right]$
                    \State append $\langle\textbf{q}\rangle \gets \left[\langle\textbf{q}\rangle\; \bar{q}_j^i\right]$
                    \State update $C_q \gets$ \Call{updateCov}{$C_q,X_q;\boldsymbol{\theta}_q$}
                    \If{$\left\{h_j^i, \rho_j^i\right\} \neq X_p$}
                        \State append $X_p \gets \left[X_p\; \left\{h_j^i, \rho_j^i\right\}\right]$
                        \State append $\langle\textbf{p}\rangle \gets \left[\langle\textbf{p}\rangle\; \bar{p}_j^i\right]$
                        \State update $C_p \gets$ \Call{updateCov}{$C_p,X_p;\boldsymbol{\theta}_p$}
                    \EndIf
                \EndProcedure
            \EndIf
        \EndFor
        \State $\boldsymbol{\rho}_{1...N}^{i+1} \gets \Delta t (1/A) \Delta\bar{\textbf{q}}_{1...N}^i/\Delta\textbf{S}_{1...N}\;\;\;\; \triangleright$ Eq.\ \eqref{continuityEq}
    \EndFor
\EndProcedure
\end{algorithmic}
\end{algorithm}

The nature of the hybrid method means the pressure needs to be evaluated before the mass flow rate, as the mass flow rate is dependent on the force, which is in turn dependent on the pressure gradient. This is problematic because, when we encounter a `new' configuration (i.e. one for which we cannot accurately predict the output), we do not want to perform two separate MD simulations for pressure and mass flow rate. We overcome this problem by ensuring the pressure prediction is always accurate, i.e. the pressure database is well represented across all of input space. This is more simple than it first sounds, because pressure is only dependent on two inputs: channel height and density. The channel heights at each subdomain are known and unchanging throughout a single case,   so to robustly map density we only need to perform one pre-simulation per subdomain. In each pre-simulation, we start from a high density of 1800 kg/m$^3$ and measure the pressure as atoms are progressively removed, down to a density of 800 kg/m$^3$. Our GP regression for pressure is therefore similar to the empirical equation of state of Borg et. al. \cite{unsteadyIMM}. However, using a GP, we avoid having to guess the functional form of this equation of state. These pre-simulations are sufficiently cheap that they do not affect our overall computational efficiency calculations.

As the macro model evolves in time, a configuration $\textbf{x}_*=\left\{h_j^i,\rho_j^i,F_j^i\right\}$ is tested to provide micro information, where $j=1,N$ is the spatial index (representing subdomain location) and $i=1,t_{\mathrm{end}}/\Delta t$ is the temporal index; $t_{\mathrm{end}}$ and $\Delta t$ are the end time and time step, respectively, for the macro model. If ${\sigma_q}_j^i > \sigma_t$, then a new MD simulation is performed with input $\textbf{x}_*$. In our MD subdomain simulations, output measurements are averaged over 40000 time-steps (see \ref{appendA}, MD simulation details) after the initial start-up phase. To minimise the start-up phase, new MD simulations are initiated with the final atomistic positions and velocities of a previously simulated configuration from our database; the input for this prior configuration $\textbf{x}_{\mathrm{in}}$ will maximise the covariance with the new test input $K(\textbf{x}_{\mathrm{in}},\textbf{x}_*)$. If there is no database entry for the channel height to be tested, new MD simulations start with the fluid atoms in a lattice, with zero mean velocity. We estimate the time for the start-up phase $t_s$ by performing a simple `pseudo MD' simulation using a 1D Navier-Stokes solver with a Navier slip condition.

After the simulation is complete, the measured mass flow rate $\bar{q}_j^i$ is appended to the vector $\langle\textbf{q}\rangle$, the input $\textbf{x}_*$ appended to the input matrix $X_q$, and the covariance matrix for the data $C_q(\left[X_q\; \textbf{x}_*\right],\left[X_q\; \textbf{x}_*\right])$ is updated. If $\left\{h_j^i,\rho_j^i\right\}$ do not exist in the pressure database, then the pressure input matrix $X_p$, output vector $\langle\textbf{p}\rangle$, and covariance matrix $C_p(\left[X_p\; \left\{h_j^i,\rho_j^i\right\}\right],\left[X_p\; \left\{h_j^i,\rho_j^i\right\}\right])$ are also updated.

\section{Results and Discussion} \label{secResults}
\noindent Without recourse to experimental results, we test the accuracy of our hybrid scheme by comparing it to a full MD simulation for the same system; this also enables us to directly quantify the computational savings of our scheme. All the full MD solutions presented here are taken from Borg et. al. \cite{unsteadyIMM}, with data points representing block averages over 2000 time-steps to reduce noise. For the majority of the results we present, the external forcing $F_{\mathrm{ext}}$ is sinusoidal with an amplitude of $F_A=0.487$ pN and a period of $T=10.8$ ns (Case C in \cite{unsteadyIMM}), i.e.
\begin{equation}
F_{\mathrm{ext}}(t) = F_A\sin\left(\frac{2\pi t}{T}\right). \label{extForcing1}
\end{equation}
This large oscillation period ensures a high degree of temporal scale separation. Let us first consider the case where we initially have an empty MD database. This could be considered the `purest' form of machine learning in our scheme, as the input ranges that need to be learned are not influenced by human intuition. While it is fairly straightforward to estimate the input ranges for our benchmark system (as explained later), this may not be true for more complex systems, with a larger number of inputs; it is therefore important to demonstrate that our scheme is sufficiently robust to accurately model the flow behaviour, even when we have no prior information. Nevertheless, setting the target uncertainty threshold $\sigma_t$ involves some subjectivity. A sensible approach is for the uncertainty threshold to be set larger than the observation noise (${\sigma_n}_q$), because it is difficult for the model to make predictions with more accuracy than the data upon which it is based\footnote{If there are multiple data points within close proximity to the test input, then the uncertainty can be lower than the observation noise.}. Noting from \S\ref{secHypers} that ${\sigma_n}_q=0.05$ ng/s, we initially choose a threshold of $\sigma_t=0.1$ ng/s. This will be referred to as Case 1.
\begin{figure}[]
\centering
\includegraphics[width=\textwidth]{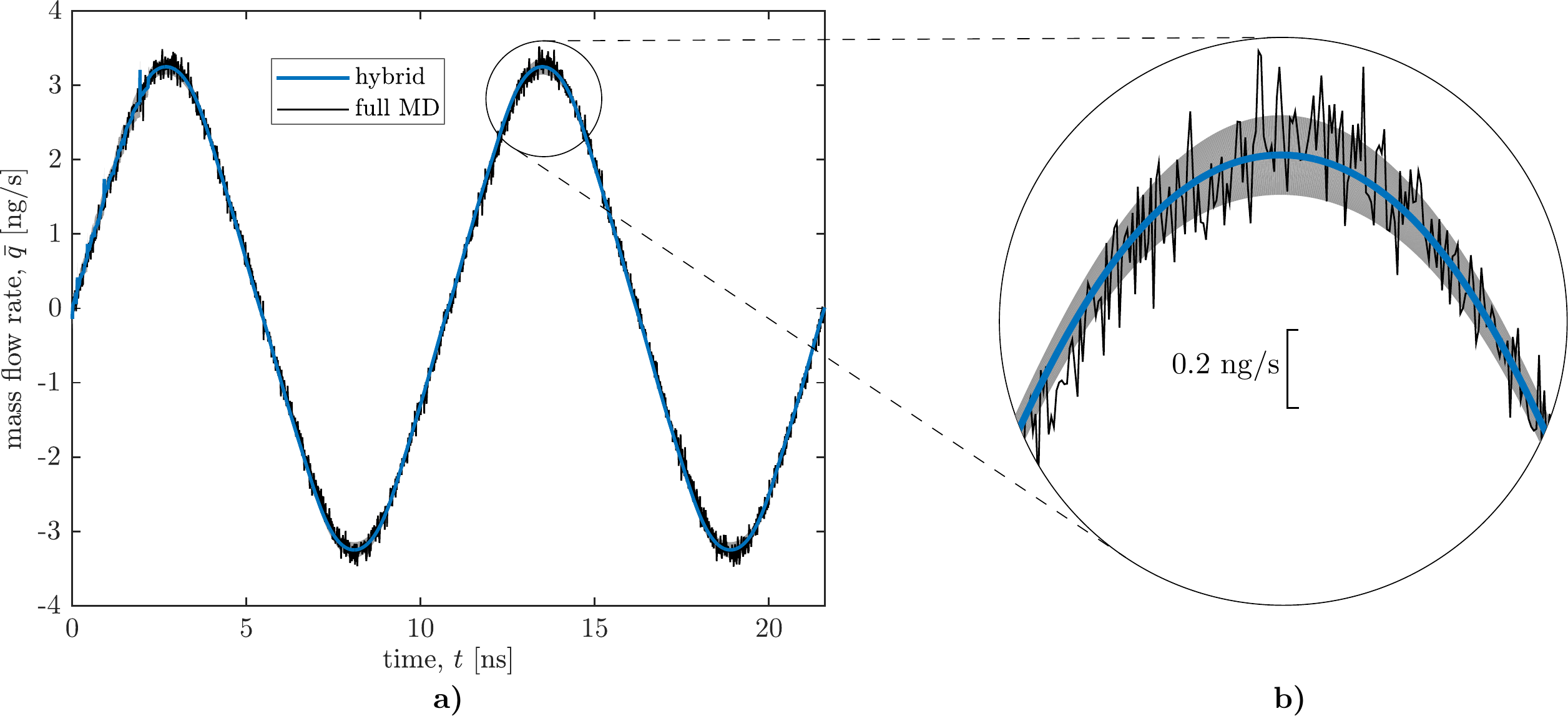}
\caption{Transient mass flow rate results for our hybrid scheme (Case 1\,---\,subdomain 1) and the full MD simulation \cite{unsteadyIMM}, showing a) the full time series, and b) a close-up to highlight the uncertainty quantification of the hybrid solution\,---\,the grey envelope represents 1.96 standard deviations of the mean, representing the 95\% confidence interval for our hybrid solution. The hybrid solution uses a tight uncertainty threshold of 0.1 ng/s and starts from an empty database.}
\label{mfrFig}
\end{figure}

The transient mass flow rate results for Case 1 are displayed in Fig.\ \ref{mfrFig}, showing excellent agreement shown between the output of our hybrid scheme and the measurements from the full MD simulation. Mass conservation means that the mass flow rate profile is approximately the same at all subdomain locations, so we present the data only for subdomain 1. As we begin from an empty database, initially our hybrid scheme must perform MD simulations with high frequency, because there is limited data upon which to base a prediction. Therefore, the hybrid solution (blue line) exhibits noise similar to that of the full MD simulation up until $t=2.7$ ns, where the external forcing function peaks. As the system geometry and external forcing function are both symmetric, after this time no `new' input configurations are experienced, and no further MD simulations need to be performed. Beyond this time, our hybrid solution near-perfectly captures the sinusoidal variation of mass flow rate, with smoothness resulting from our covariance prior. Figure \ref{mfrFig}b shows a close-up of the second mass flow rate peak and highlights that the uncertainty of our hybrid solution (grey region, representing 95\% confidence bounds) is smaller than the noise in the full MD simulation. This is because the properties in the full MD simulation are transient over time-averaging for each data point, whereas our MD subdomain simulations are steady-state. The uncertainty is larger at the extremes of mass flow rate because these configurations exhibit the most extreme force and density inputs, so the model must extrapolate beyond its existing database.

While building the database during the first 2.7 ns, the model continually has to extrapolate for mass flow rate predictions (as the increasingly large forces and densities render existing data less relevant to the current conditions), until the uncertainty threshold is reached and a new MD simulation is performed. As the first mass flow rate peak approaches, there will be a final MD simulation, beyond which all configurations are sufficiently close to those in the database that the uncertainty of future predictions is always below the threshold. This means that at all subsequent peaks, the model must extrapolate from the database, whereas, for all other configurations, the model will interpolate within the database. One potential benefit of our machine-learning approach is the ease in which the effect that uncertainty in the external forcing function has on the mass flow rate solution can be quantified. Using pure MD simulations or traditional hybrid methods, it would be an extremely cumbersome task to model a wide range of forcing functions; however, with our scheme, once the database is built, new cases can be performed incredibly quickly (as shown in \S\ref{secBuilding}).

\begin{figure}[]
\centering
\includegraphics[width=\textwidth]{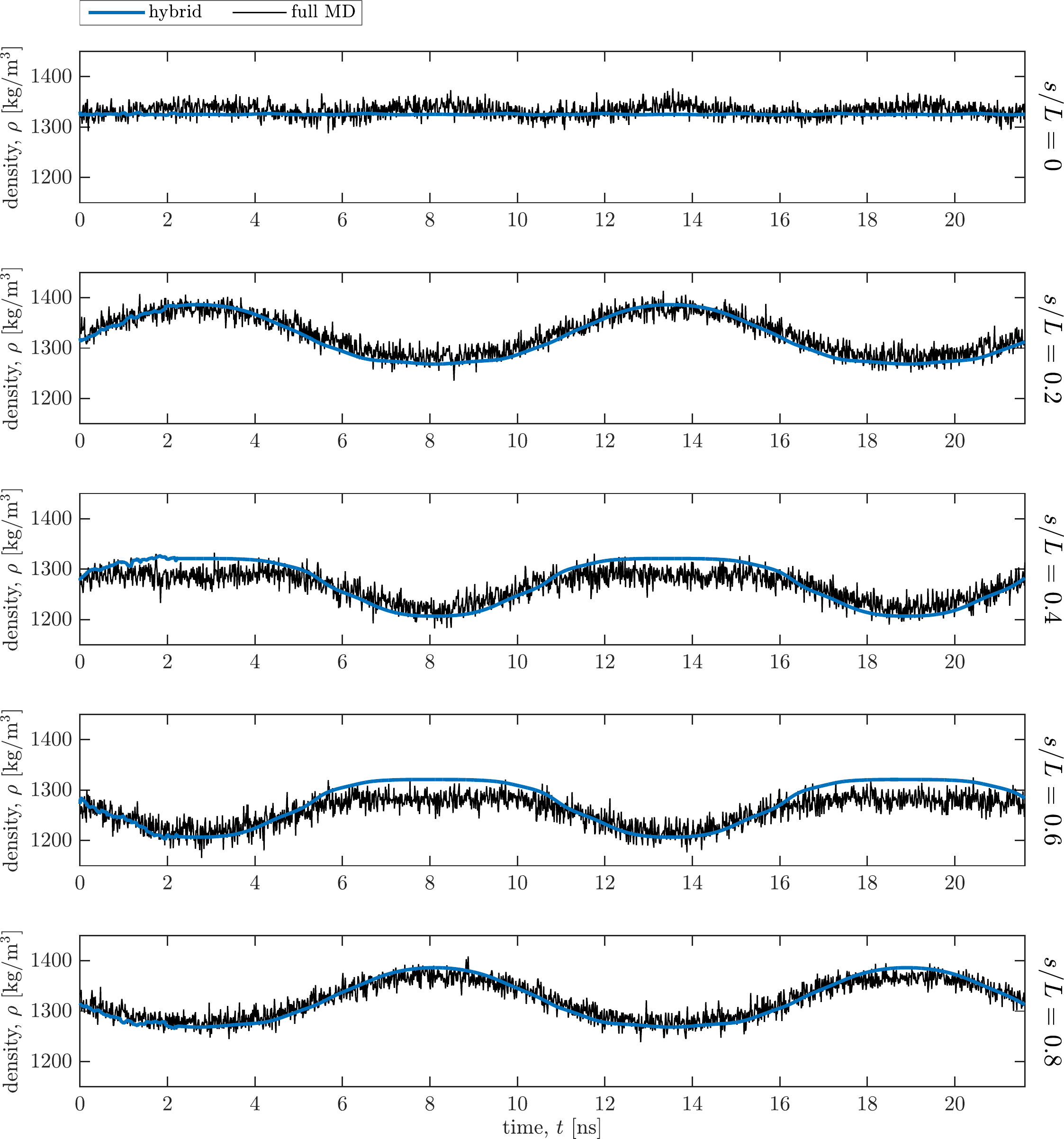}
\caption{Transient density results for our hybrid scheme (Case 1) and the full MD simulation, calculated at each of the five subdomains. The hybrid solution uses a tight uncertainty threshold of 0.1 ng/s and starts from an empty database.}
\label{densityFig}
\end{figure}
Figure \ref{densityFig} displays the transient density profiles for Case 1, calculated at each of the five subdomains. Our hybrid solution shows good agreement with the measurements from the full MD simulation, although it fails to produce the subtle oscillations in subdomain \#1, and occasionally overestimates the density in subdomains \#3 and \#4. However, despite overestimating the density, our hybrid solution successfully captures the non-harmonic profile in these sections. The discrepancies our solution displays mirrors those in the solutions of Borg et. al., so are likely due to a common inaccuracy in the multiscaling, rather than an error in our GP regression model. We note that since density is an input to rather than an output of the regression model there is no simple method for uncertainty quantification.

\subsection{The effect of the uncertainty threshold} \label{secResultsThresh}
\begin{figure}[]
\centering
\includegraphics[width=\textwidth]{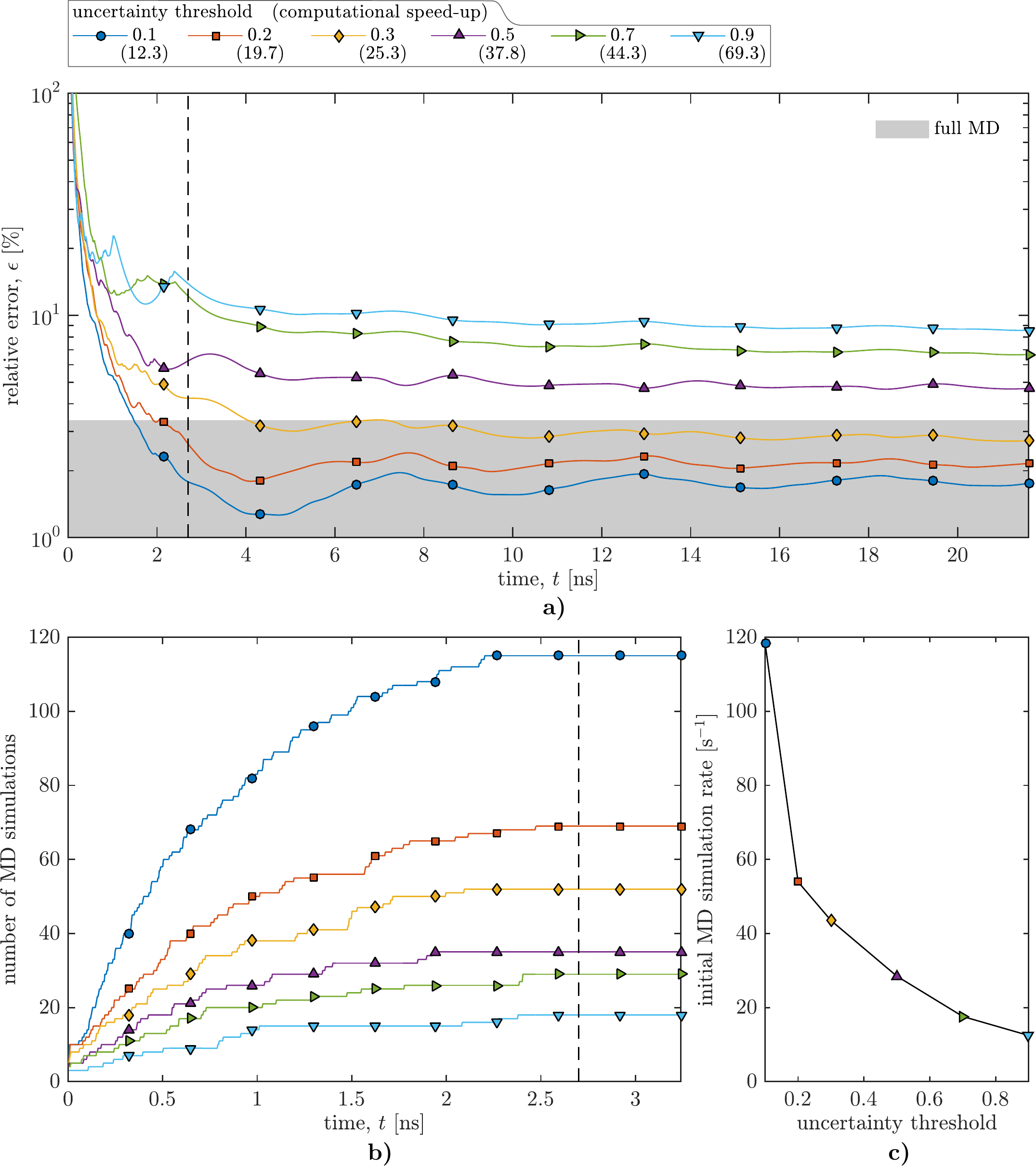}
\caption{The influence of the uncertainty threshold $\sigma_t$ on the hybrid solution for mass flow rate (starting from an empty database): a) the transient relative error to the smooth full MD solution (the grey area shows the error due to noise in the full MD solution); b) the cumulative number of MD simulations; c) the initial MD simulation rate. The vertical dashed line in a) and b) denotes time at the first peak in the external forcing function, after which no MD simulations are performed. The data is from Cases T1 ($\sigma_t = 0.1$ ng/s) to T6 ($\sigma_t = 0.9$ ng/s), and the computational speed-up over the full MD solution for each case is displayed in parentheses in the legend.}
\label{threshFig}
\end{figure}
\noindent Cases T1\,--\,T6 demonstrate the effect of the choice of uncertainty threshold on our hybrid solution. Case T1 is the same as Case 1, and the threshold increases up to 0.9 ng/s for Case T6. To isolate the effect of the threshold, Cases T1\,--\,T6 all start with an empty database. Figure \ref{threshFig} confirms our intuition that the choice of threshold in our hybrid scheme is a trade-off between the accuracy and computational efficiency of its solution. As the threshold is raised, MD simulations are performed less frequently, so the accuracy of our hybrid solution drops, but the speed of computation increases. The signal standard deviation is ${\sigma_f}_q=1$ ng/s (see \S\ref{secHypers}), so when the threshold $\sigma_t>1$ ng/s, MD simulations will never be performed, even when starting from an empty database. In this instance, the mass flow rate in each subdomain is predicted to be zero for the entire time series, as this is the assumed mean (see \S\ref{secML}).

Figure \ref{threshFig}a shows the total mass flow rate error $\epsilon$ for each case, relative to the error obtained if the database remains empty (which represents the maximum possible discrepancy), defined by
\begin{equation}
\epsilon_t = \frac{\sum_{i=1}^t\lvert\bar{q}_i - Q_i\rvert)}{\sum_{i=1}^t \lvert Q_i\rvert} \times 100\%, \label{error}
\end{equation}
where $\bar{q}_i$ is the mass flow rate, which is either measured by a MD subdomain simulation (when ${\sigma_q}_i>\sigma_t$), or predicted by our regression model (when ${\sigma_q}_i \leq \sigma_t$) at time index $i$ (averaged across all subdomains), and $Q_i$ is the mass flow rate from the full MD simulation. To obtain a meaningful error, the noise from the full MD solution is filtered by performing GP regression over the raw data, using a periodic kernel with time as the single input (see \ref{appendB} for details). At first, the errors are greater than 100\%. This is because the weak external forcing generates a near-zero mass flow rate, so the measurements from MD simulations (which are performed frequently at the start) are dominated by thermal fluctuations and masked by noise. Somewhat counterintuitively, the errors continue to drop after the final MD simulation is performed; this is because the weight of the errors incurred early in the trajectory diminishes as the hybrid solution becomes more accurate after the first peak in the forcing function. To obtain context for the error magnitudes, we also use equation \eqref{error} to calculate the `error' in the raw full MD simulation data (grey area); for clarity, the calculation at time $t=21.6$ ns is shown for the entire time series. Our results suggest that up to Case T3, where $\sigma_t=0.3$ ng/s, the error in our hybrid solution is comparable to the error in the full MD simulation due to noise.\footnote{It would be wrong to say that our hybrid solution is more accurate than the full MD solution.} Overall the error increases with increasing threshold, as is expected. 

Figures \ref{threshFig}b and c demonstrate how the computational speed of our scheme varies with the uncertainty threshold. Figure \ref{threshFig}b shows the cumulative number of MD simulations, which is proportional to the efficiency of the solution. As expected, the trend is that the lower the threshold, the more MD simulations must be performed. In all cases no further MD simulations are required after $t=2.7$ ns. The computational speed-up, given in brackets in the legend of Fig.\ \ref{threshFig}, is calculated by
\begin{equation}
\mathrm{speed\mbox{-}up} = \frac{t_{\mathrm{end}}N_{a_f}}{\bar{t}_{\mathrm{sim}}N_{\mathrm{sim}}\bar{N}_{a_h}}, \label{speedup}
\end{equation}
where $N_{a_f}$ is the number of atoms in the full MD simulation, $\bar{t}_{\mathrm{sim}}$ is the average time-steps performed in a single MD subdomain simulation, $N_{\mathrm{sim}}$ is the number of MD subdomain simulations performed for the hybrid solution, and $\bar{N}_{a_h}$ is the average number of atoms in each of those MD simulations. For the tightest threshold (Case T1), our hybrid solution provides a modest speed-up over the full MD simulation of 12.3$\times$; this rises to 69.3$\times$ for the loosest threshold (Case T6), however the relative error of this solution is 2.5$\times$ that of the full MD simulation. All cases (T1\,--\,T6) all show logarithmic growth for the total number of MD simulations, with respect to time\,---\, i.e. the frequency of MD simulations decreases as the database becomes larger, and the predictions become more accurate. 

Figure \ref{threshFig}c shows how the initial required MD simulation rate varies with the uncertainty threshold. This can be considered a quantification of the `novelty' of the solution\,---\,i.e. the rate of identifying new configurations that are dissimilar to those on which the model has been previously trained. We calculate this novelty by assuming the MD simulation rate is approximately constant over the first 0.27 ns.\footnote{For the variable threshold run, the $x\mathrm{-axis}$ in Fig.\ \ref{threshFig}c is taken to be the average threshold over the first 0.27 ns.} The results confirm that lower thresholds produce a higher novelty as they require more certainty for a configuration to be recognised as being similar to those in the existing database.

\subsection{The effect of the initial database} \label{secResultsData}
\noindent As previously acknowledged, prior to running our hybrid scheme, we can make some good estimations for the range of input values in our example case. For the particular system of interest studied here, we know that the channel height will always be equal to either $h_1$, $h_2$, or $h_3$ (see Fig.\ \ref{introFig} for reference; it is likely that the magnitude of the total force $\lvert F\rvert$ will range from zero to at least the magnitude of the external force (0.487 pN); and from knowledge of nanofluidic MD simulations, we conservatively estimate that density will vary between 1120 kg/m$^{\mathrm{3}}$ and 1480 kg/m$^{\mathrm{3}}$\,---\,approximately $\pm 180$ kg/m$^{\mathrm{3}}$ from the initial values. Simulations are performed at evenly-spaced increments across these ranges to create an initial database to train the model. Subsequently, `new' configurations are encountered less frequently, and more predictions are made by interpolating between existing database entries, rather than extrapolating. This means predictions are made with more certainty, and the total number of simulations is likely to be reduced.

\begin{table}[h]
\caption{Initial databases for Cases D1\,--\,D5. See Fig.\ \ref{introFig} for channel height references.}
\vskip 2pt
\centering
\begin{tabular}{c c c}    \toprule
Case & Channel heights & Initial database size \\\midrule
D1 & $-$ & 0 \\
D2 & $h_1$ & 15 \\
D3 & $h_3$ & 16 \\
D4 & $h_1, h_3$ & 31 \\
D5 & $h_1, h_2, h_3$ & 47 \\
\hline
\end{tabular}
\label{preDatabases}
\end{table}
The initial databases for cases D1\,--\,D5 are outlined in Table \ref{preDatabases}, where Case D1 is the same as Case 1\,---\,i.e. the initial database is empty. The other four databases are trained using one or more channel height and, for each height, four different densities and forces, evenly spaced across the estimated ranges above. The exception is for channel height $h_1$, the largest force (0.487 pN) applied to the lowest density fluid (1120 kg/m$^{\mathrm{3}}$) produces a shear rate beyond the critical limit \cite{critShear}, so the mass flow rate does not converge; therefore, this data entry is removed from the initial databases. For all cases, the uncertainty threshold is set at $\sigma_t=0.1$ ng/s.

\begin{figure}[]
\centering
\includegraphics[width=\textwidth]{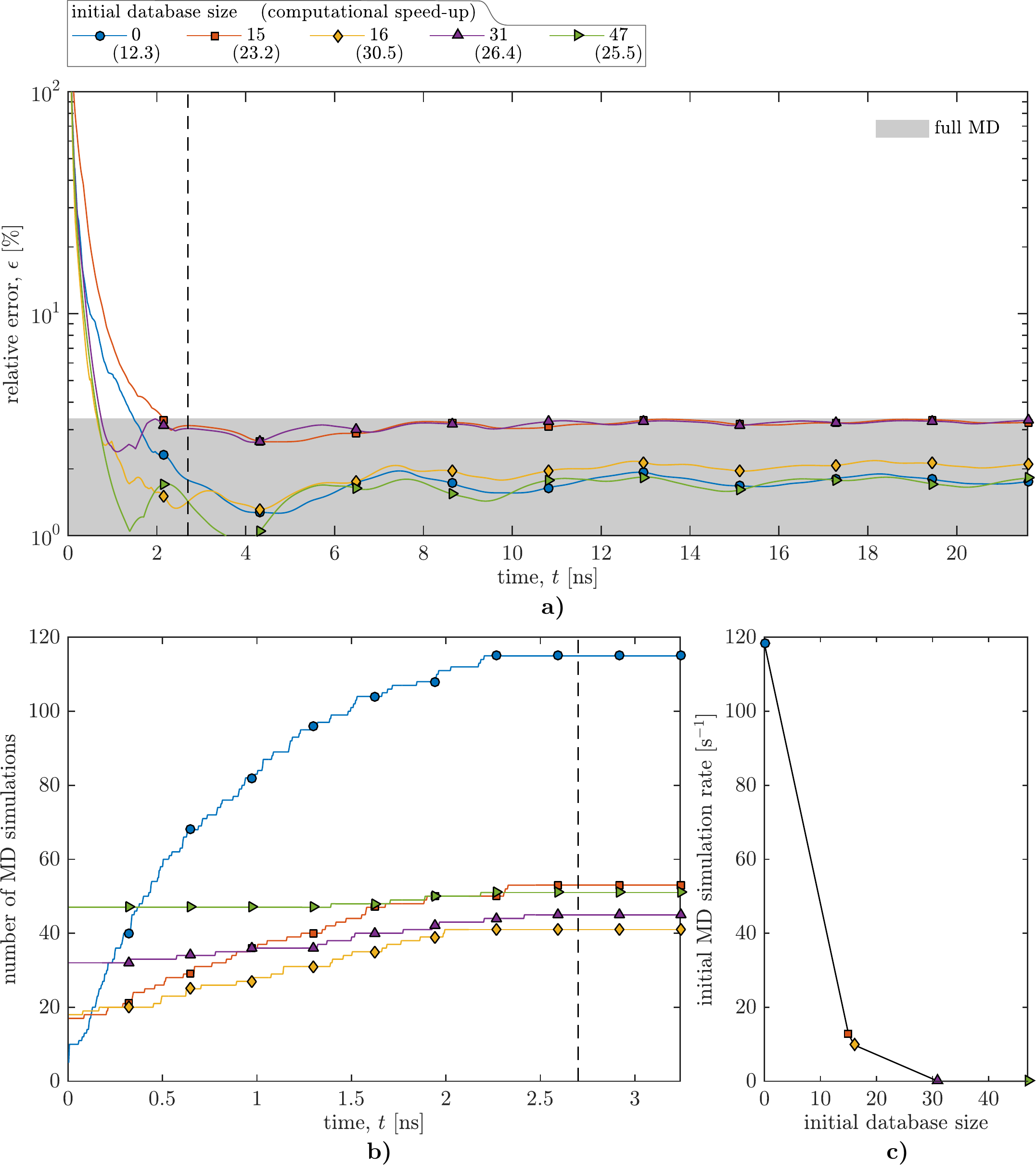}
\caption{The influence of the initial database on the hybrid solution for mass flow rate (with an uncertainty threshold of 0.1 ng/s): a) the transient relative error to the smooth full MD solution (the grey area shows the error due to noise in the full MD solution); b) the cumulative number of MD simulations required; c) the initial MD simulation rate. The vertical dashed line in a) and b) denotes time for the first peak in the external forcing function, after which no MD simulations are performed. The data is from Cases D1\,--\,D5 (see Table \ref{preDatabases}, and the computational speed-up for each case is displayed in parentheses in the legend.}
\label{priorFig}
\end{figure}
The results for Cases D1\,--\,D5 are presented in Fig.~\ref{priorFig}. Figure \ref{priorFig}a demonstrates that the size of the initial database has a negligible effect on the overall accuracy of the method, with all cases exhibiting errors comparable to the noise in the full MD simulation. This result is expected, because the uncertainty produced by extrapolating beyond a smaller initial database is countered by more frequently performing MD simulations.

Figure \ref{priorFig}b shows that the total number of MD simulations decreases when the initial database is not empty, because more predictions are made through interpolation and `new' configurations are not found so frequently at the start of the time series. This effect is demonstrated in Fig.~\ref{priorFig}c, where the novelty of the cases decreases as the initial database increases, reaching zero for the largest database in Case D5. However, the total number of MD simulations performed does not continue to fall as the initial database grows. For larger initial databases, sometimes redundant information is gathered which could have been inferred from neighbouring configurations; this excessive training outweighs the benefit of the increased confidence in subsequent predictions. This effect is demonstrated in the results for Case D5, where increasing the initial database to include simulations with channel height $h_2$ only negligibly reduces the number of `on-the-fly' simulations performed compared to Case D4. 

Another example of redundant information is the discrepancy between the results for Cases D2 and D3, Despite the model training on only one extra configuration for the latter case. This is due to the geometry of the case: a much larger force is required near the throat of the channel (subdomain \#3) than at the inlet/outlet (subdomain \#1) to generate equal mass flow rates.\footnote{It is difficult to provide similar analysis for the density ranges, because at the nanoscale, viscosity varies rapidly with density, so greater density does not necessarily produce a larger mass flow rate.} As such, the local pressure gradient always acts in the opposite direction to the external force at the inlet/outlet of the channel, so the peak external force is never experienced in subdomain \#1, and training on it for initial database provides little information. Conversely, all of the information is used when the initial database is formed using subdomain \#3. Fewer MD simulations corresponds to an increase in computational efficiency\,---\,Case D3 is 30.5$\times$ faster than the full MD simulation while maintaining the same level of accuracy.

\subsection{Building on an existing database} \label{secBuilding}
\noindent As we have already demonstrated with previous results, one important advantage of our hybrid scheme is that it enables information to be reused so predictions can be made with more certainty from a continually-growing database. So far, this information has been reused by increasing the confidence of our mass flow rate predictions for configurations later in the time series, within the same case. However, we can go further. For example, suppose having completed the hybrid simulation, we decided that we are really interested in a flow feature occurring at $s=6.8$ nm (halfway between subdomains \#1 and \#2). Our previous options would have been to run the expensive full MD simulation to ensure every flow feature is captured, or to perform a new hybrid simulation using different subdomain locations; both of which are computationally wasteful. However, with our machine-learning-accelerated hybrid scheme, we can simply create a new case which has more subdomains, with the model already trained on the database that we generated in the previous case.

\begin{figure}[]
\centering
\includegraphics[width=\textwidth]{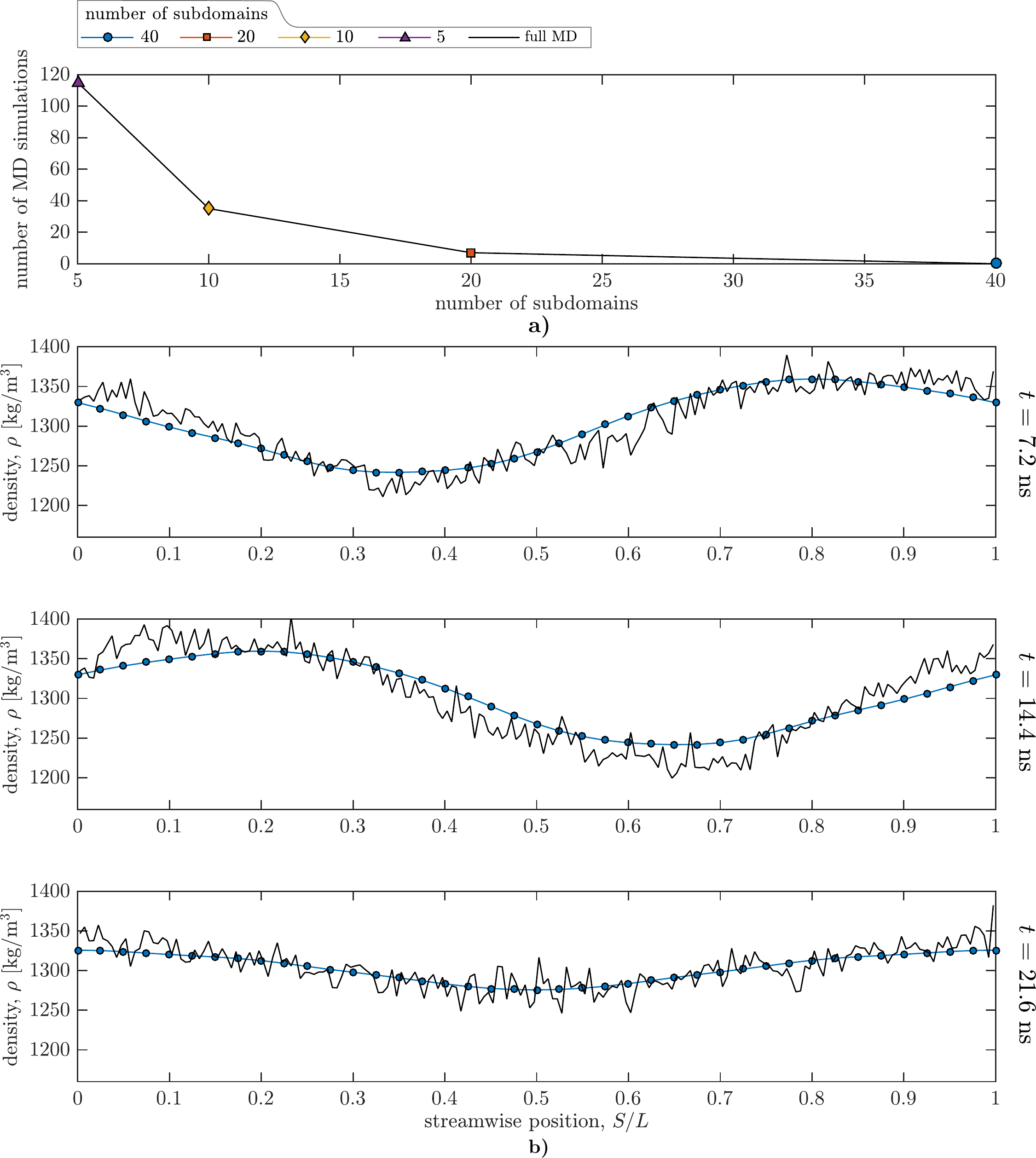}
\caption{The influence of the number of subdomains on the hybrid solution for density (with an uncertainty threshold of 0.1 ng/s): a) the number of new MD simulations; and b) streamwise density profiles for the full MD simulation and the hybrid solution using 40 subdomains at $t = 7.2$ ns, $t=14.4$ ns, and $t=21.6$ ns. The data is from Cases S1\,--\,S4.}
\label{subsFig}
\end{figure}

In Cases S1\,--\,S4, we demonstrate how this approach can be used to continually add subdomains and refine the streamwise density profile. Case S1 is the same as case 1; in each subsequent case, the number of subdomains is doubled (with the new subdomain locations bisecting the old subdomain locations), and the initial database is that generated at the end of the previous case.\footnote{Note, extra pre-simulations for the pressure database are not required for the new subdomain heights, as the inference is already sufficiently accurate.} For all cases, the uncertainty threshold is $\sigma_t=0.1$ ng/s. As the spacing between adjacent subdomains decreases, the relevance of data measured at neighbouring subdomains increases, and successively fewer MD simulations are performed, as shown in Fig.\ \ref{subsFig}a. Using 40 subdomains, no new MD simulations are required at all. In addition, the streamwise density profiles calculated using five subdomains are inaccurate, because central differences are used to model spatial gradients, which assumes the variation between adjacent subdomains is linear. However, as the spacing between subdomains decreases, this linear assumption becomes more accurate. Figures \ref{subsFig}b-d show how the streamwise density profiles for case S4 (40 subdomains) compares to the profile measured by the full MD simulation, at three snapshots in time. Our results show good qualitative agreement with the noisy MD data.

\begin{figure}[]
\centering
\includegraphics[width=\textwidth]{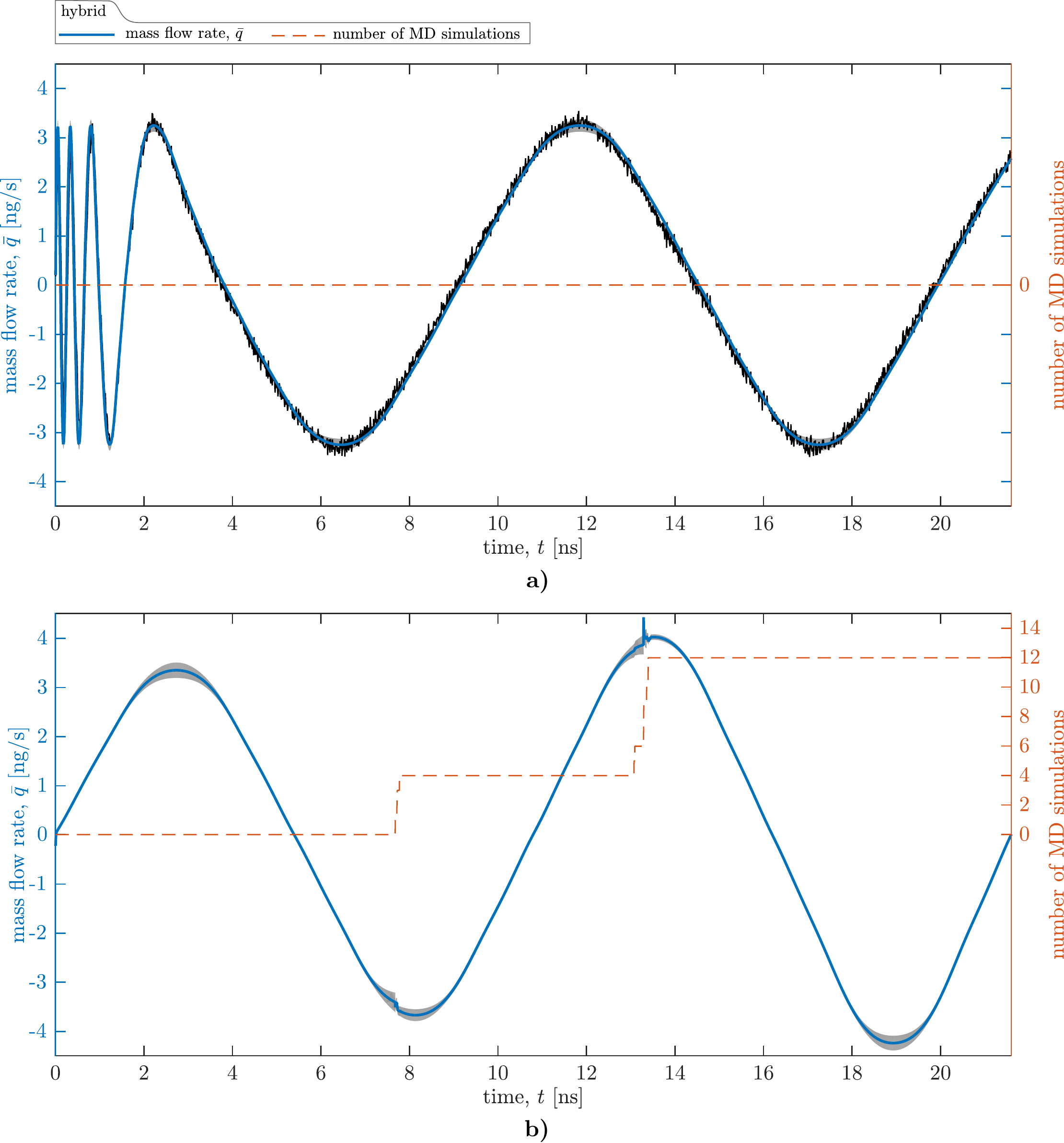}
\caption{Transient mass flow rate results (left $y\mathrm{-axis}$) for the hybrid solution of a) Case F1 (variable frequency external force), and b) Case F2 (variable amplitude external force). For Case F1, the result from the full MD simulation is also plotted. In both cases, the hybrid solution uses an uncertainty threshold of 0.1 ng/s, and an initial database generated at the end of Case 1. The cumulative number of new MD simulations is also plotted (right $y\mathrm{-axis}$).}
\label{forceFig}
\end{figure}
Another example of building on an existing database is evaluating the response to different external forcing functions in the same geometry. Again, without the aid of machine learning, this would require performing an entirely new hybrid or full MD simulation. Using the database generated at the end of Case 1 as our initial database, we perform two new cases: Case F1, with a variable-frequency external force; and Case F2, with a variable-amplitude external force. In Case F1, the oscillation period of the force starts from $0.22$ ns and gradually increases to $10.8$ ns. \cite{unsteadyIMM}; the amplitude is 0.487 pN as case 1. The left $y\mathrm{-axis}$ of Fig.\ \ref{forceFig}a shows the transient mass flow rate results for our hybrid solution and the full MD solution (Case D in Borg et. al. \cite{unsteadyIMM}), and the right $y\mathrm{-axis}$ shows the number of new MD simulations performed. Once again, our solution exhibits strong qualitative agreement with the full MD simulation, and the computational speed-up is effectively infinite\footnote{Of course, the speed-up is not actually infinite, but the computational effort to run the surrounding algorithm is neglible compared to an MD simulation.} as no new MD configurations are performed. In Case F2, the amplitude of the force starts from $0.487$ pN and gradually increases to $0.609$ pN;\footnote{A much large force would cause the critical shear limit to be exceeded.} the oscillation period is a constant $10.8$ ns as in case 1. In Fig.\ \ref{forceFig}b, we show our hybrid solution for the transient mass flow rate profile (left $y\mathrm{-axis}$) and the number of new MD simulations performed (right $y\mathrm{-axis}$). There is no full MD simulation data for this forcing function, so we present our results in part as a challenge to others. The results appear consistent with our previous findings: the mass flow rate profile peaks at the peaks of the forcing function, and the magnitude of the fourth peak is approximately 1.25$\times$ the magnitude of the fourth peak in Case 1\,---\,mirroring the variation of force amplitude. New MD simulations are only required as the mass flow rate (and therefore forcing function) approach peaks, where the GP model has to extrapolate beyond the existing force entries in the database. The lack of smoothness in our hybrid solution, where new MD simulations are required, can be remedied by re-running the case with the final frozen database, at a negligible cost.  

\section{Conclusion} \label{secConc}
We have presented an enhancement to existing hybrid methods in fluid dynamics that uses on-the-fly machine learning techniques to predict the output of new system configurations, without having to continually perform atomistic simulations. This procedure enables information to be reused multiple times, drastically increasing the computational efficiency, while also providing the potential for simple uncertainty quantification.

We compare our new scheme to full molecular dynamics (MD) simulations and find strong agreement, with errors within the range of thermal noise when a tight uncertainty threshold is set. As this threshold is raised, the error increases to up to 2.5$\times$ the MD simulation noise; however, the computational speed-up over a full MD simulation also increases. When starting from an empty database, raising the threshold from 0.1 ng/s to 0.9 ng/s increases speed-up from 12.3$\times$ to 69.3$\times$. Thus, the choice of threshold is a trade-off between the required accuracy and computational efficiency.

We demonstrate the computational benefit of creating an initial database to train our predictive model, by estimating the expected range of inputs. This enables more predictions to be made via interpolation between data points, which provides less uncertainty than extrapolation leading to a requirement for fewer MD simulations to be performed `on-the-fly'. While maintaining the same level of accuracy, starting with a modest initial database covering just a few data points can result in a speed-up of 30.5$\times$ the full MD simulation. Finally, we show how existing databases can be built upon (while never needing to be fully complete) to rapidly obtain high resolution hybrid solutions\,---\,i.e. cheaply add more subdomains at locations of interest\,---\,or to model different flow fields effectively instantly\,---\,i.e. no new MD simulations are required.

\section*{Acknowledgements}
We would like to thank Matthew Borg for providing the data for the full MD simulations. This work is financially supported in the UK by EPSRC Programme Grants EP/I011927/1 and EP/N016602/1, and EPSRC grants EP/K038664/1 and EP/L027682/1. The computing facilities were provided by the Centre for Scientific Computing of the University of Warwick with support from the Science Research Investment Fund.

\appendix
\section{MD parameters} \label{appendA}
\noindent Non-equilibrium MD simulations of dense fluid argon are performed using mdFoam \cite{OF1,OF2}\,---\,an in-house solver developed in OpenFOAM. Atoms are modelled as hard spheres which interact using shifted Lennard-Jones (LJ) pair potentials, with wall atoms frozen in place. Our MD parameters are necessarily identical to those used by Borg et. al. \cite{unsteadyIMM} for the full MD simulations. The LJ potential
\begin{equation}
U_{LJ}(r_{ij}) = 4\epsilon\left[\left(\frac{\sigma}{r_{ij}}\right)^{12} - \left(\frac{\sigma}{r_{ij}}\right)^{6}\right], \label{LJ}
\end{equation}
has a cut-off radius of 1.36 nm; for the fluid-fluid interactions, the LJ characteristic length and energy are $\sigma_{f-f}=0.34$ nm and $\epsilon_{f-f}=1.65678 \times 10^{-21}$ J, respectively; for the wall-fluid interactions, these parameters are $\sigma_{w-f}=0.255$ nm and $\epsilon_{w-f}=0.33 \times 10^{-21}$ J, respectively. The mass density of the wall atoms is $6.809 \times 10^{3}$ kg/$\mathrm{m^3}$, and the mass of a single atom is $m=6.6904 \times 10^{-26}$ kg. Fluid atom dynamics are described by Newton's laws of motion, which are numerically integrated using the Velocity Verlet method \cite{Swope}, with a time-step of 5.4 fs. The excess heat generated by applying an external force is removed by modifying the velocities in the $z$-direction (see Fig.\ \ref{introFig}) using a Berendsen thermostat; this ensures a thermally homogeneous system, maintained at a temperature of 292.8 K. All of our simulations are periodic in the $s$- and $z$-directions.

\section{Gaussian process regression of the full MD simulation data} \label{appendB}

\noindent To obtain error estimates for our hybrid solution, the full MD mass flow rate profile is modelled using a Gaussian process with a single input: time $t$. As the underlying profile is sinusoidal, we use a periodic kernel, i.e.
\begin{equation}
K(t,t') = \sigma_f^2 \exp\left(-\frac{2\sin^2\left(\pi\lvert t-t'\rvert/T\right)}{\ell^2}\right), \label{periodicKernel}
\end{equation}
where $T=10.8$ ns is the period of oscillation; the covariance function is still calculated using equation \eqref{covarianceEq}. For consistency with our hybrid scheme, we choose the signal standard deviation $\sigma_f=1$ ng/s. We anticipate the noise standard deviation to be larger than in our quasi-steady subdomain simulations, due to transience of properties during the averaging period in the full MD simulation. Therefore, we optimise for the length scale and the noise standard deviation by maximising the log marginal likelihood of the raw full MD data (up to the first peak), as outlined in \S\ref{secHypers}, and find that $\ell=1$ and $\sigma_n=0.1$ ng/s (twice that of our MD subdomain simulations). Gaussian process regression is performed over the entire time series using equations \eqref{predMeanEq} and \eqref{predVarEq}, with the mean mass flow rate assumed to be zero.

\section*{\refname}
\bibliographystyle{elsarticle-num}
\bibliography{stephensonKermodeLockerby}
\end{document}